\documentclass[11pt]{article}

\usepackage[utf8]{inputenc}
\usepackage[T1]{fontenc}
\usepackage{lmodern}
\usepackage[margin=1in]{geometry}
\usepackage{microtype}

\usepackage{amsmath,amssymb}
\usepackage{booktabs}
\usepackage{longtable}
\usepackage{graphicx}
\usepackage{float}
\usepackage{siunitx}
\usepackage[version=4]{mhchem}
\usepackage{algorithm}
\usepackage{algpseudocode}
\usepackage{enumitem}
\usepackage{xcolor}

\usepackage[round,authoryear]{natbib}
\let\cite\citep
\usepackage{hyperref}
\hypersetup{
  colorlinks=true,
  linkcolor=black,
  citecolor=black,
  urlcolor=black,
  pdftitle={WeDesign: Generative AI–Facilitated Community Consultations for Urban Public Space Design},
  pdfauthor={Rashid Mushkani, Hugo Berard, Shin Koseki}
}

\graphicspath{{./}}
\newcommand{\Description}[2][]{}

\newcommand{\keywords}[1]{\noindent\textbf{Keywords:} #1}

\title{WeDesign: Generative AI–Facilitated Community Consultations for Urban Public Space Design}

\author{Rashid Mushkani\textsuperscript{1,2} \quad
        Hugo Berard\textsuperscript{1} \quad
        Shin Koseki\textsuperscript{1,2}}
\date{\textsuperscript{1}Universit\'e de Montr\'eal \quad
      \textsuperscript{2}Mila -- Quebec AI Institute\\[0.25em]
      \vspace{-0.5em}
}

\begin{document}
\maketitle

\begin{abstract}
Community consultations are central to urban planning but often constrained by resources, language, and power imbalances. We study how generative text‑to‑image models can support more equitable dialogue. Using WeDesign, a custom platform integrating Stable Diffusion XL, we ran a half‑day workshop in Montréal with five mixed focus groups (residents, designers, AI specialists) and conducted six interviews with urban‑planning professionals. Real‑time images stimulated creativity, helped participants exchange ideas, and supported iterative refinement. Participants nevertheless reported limits in depicting reduced‑mobility needs, local architectural cues, and bilingual prompting; many outputs felt polished yet de‑contextualized. Participants and experts recommended open‑source tooling with region‑specific in‑painting, multilingual support, group voting, and preference sliders, alongside structured facilitation. We discuss how generative AI can broaden participation while requiring safeguards and expectations management, contributing empirical evidence on opportunities and limits of text‑to‑image AI in participatory urban design.
\end{abstract}


% ================================================================
\section{Introduction}
Public participation has become a cornerstone of contemporary urban planning, grounded in the recognition that local residents possess essential experiential knowledge of their surroundings \cite{brabham2009crowdsourcing,fischer2000citizens,foroughi2023public}. Scholars and practitioners have emphasized the need for inclusive processes, arguing that inadequate community representation risks perpetuating spatial inequities and undermining trust in governance \cite{arnstein1969ladder,davies2020online,pissourios2014topdown}. Although these concerns have prompted various consultation formats—from town halls to online forums—challenges such as language barriers, technological resource constraints, and power imbalances often impede meaningful engagement \cite{arana2021citizen,barendregt2024public,davies2020online}.

Over the last decade, new digital tools have emerged to reshape participatory urban design \cite{davies2020online}. Interactive platforms, virtual simulations, and online mapping applications have expanded avenues for engagement, yet persistent issues remain \cite{barendregt2024public}. Biases may become embedded in digital environments, digital literacy gaps can exclude certain demographics, and advanced visualization software often proves too expensive or complex for widespread adoption \cite{arana2021citizen,li2020analysis}. Concurrently, progress in generative artificial intelligence has led to increasingly sophisticated text-to-image models \cite{bendel2023image,rombach2022latent,vonbrackel2024equipping}. Systems such as Stable Diffusion, DALL·E~3, and Midjourney can create detailed images from concise prompts \cite{midjourney2022website,openai2023dalle3,rombach2022latent}, and some researchers suggest that these models could broaden participation in design processes by lowering technical barriers and enabling more immediate visual communication \cite{kucevic2024promptathon}.

This study examines the application of a generative text-to-image approach in Montreal to better understand how real-time image rendering might influence community consultations for urban public space design. Specifically, it explores how residents and local experts of diverse ages, cultural backgrounds, and abilities engaged with Stable Diffusion XL through WeDesign, a platform developed for the workshops to facilitate prompt-based visualization and iterative design refinement. Three key research questions guide the analysis:

\begin{enumerate}[leftmargin=*]
    \item \textbf{Engagement and creativity:} How does an AI-based visualization process affect participant engagement, creativity, and sense of ownership in community consultations?
    \item \textbf{Inclusivity and local features:} To what extent can generative models represent the aspirations of diverse populations and localized cultural elements?
    \item \textbf{Operational and ethical constraints:} What operational or ethical issues arise, including bilingual contexts and biases in training data, when applying text-to-image tools in urban planning?
\end{enumerate}

Drawing on an in-person workshop involving 29 participants organized into five groups and semi-structured interviews with six urban experts, this paper examines how generative AI interacts with traditional consultation processes. The findings highlight the need for better tailored platforms and indicate that, although real-time image generation can stimulate creativity and facilitate communication, limitations persist in adequately representing the needs of marginalized groups and accommodating multilingual contexts. These observations contribute to broader discussions on the responsible deployment of digital innovation to promote equitable participation in urban planning. 

\paragraph{Contributions.}
This paper makes three contributions to HCI and participatory urban design:
\begin{enumerate}[leftmargin=*]
  \item An empirical account of \emph{in-situ}, bilingual, AI-mediated community consultations using Stable Diffusion XL within a purpose-built platform, \emph{WeDesign}.
  \item A set of design implications for equitable facilitation with text-to-image models, including region-specific in-painting, multilingual prompt mediation, group preference aggregation, and parameter steering via non-technical controls.
  \item A critical analysis of representational gaps---where tokenistic depictions and generic urban archetypes obscure accessibility, heritage, and feasibility---together with facilitation strategies to mitigate expectation inflation.
\end{enumerate}

The following sections review the relevant literature, outline the study’s methodology, present the findings, and conclude with insights for integrating text-to-image AI into consultation frameworks through improved platform design and careful facilitation to address biases and linguistic barriers.

\section{Literature Review}

\subsection{Participatory urban design}
Participatory urban design encompasses strategies that incorporate local perspectives into decision-making processes \cite{arnstein1969ladder,brabham2009crowdsourcing,pissourios2014topdown}. City governments and planning agencies often employ workshops, charrettes, or online consultations to solicit community feedback. This approach is grounded in the rationale that local inhabitants possess firsthand knowledge of a neighborhood’s cultural, social, and infrastructural needs \cite{fischer2000citizens}. Yet even well-structured participatory formats can face limitations such as low attendance, lack of representativeness, and difficulties in effectively communicating complex spatial proposals \cite{cooke2001participation,davies2020online,fors2021striving,li2020analysis}.

One area of concern is the “technocratic overshadowing” \cite{li2020analysis} that can occur when professional designers or municipal staff rely on specialized software and jargon-heavy presentations. Residents may struggle to visualize proposals or articulate alternative ideas, leading to consultations that appear formal but fail to integrate substantive input \cite{cooke2001participation,davies2020online,fors2021striving}. Initiatives that place community members in a more central position—either through iterative co-creation or accessible design tools—have been shown to promote deeper engagement \cite{brabham2009crowdsourcing,fors2021striving,rinaldi2020codesign}.

\subsection{Digital platforms}
The proliferation of digital engagement platforms has introduced new opportunities and dilemmas \cite{davies2020online,lau2024democratizing}. Tools such as Pol.is, Decidim, Decide Madrid, and Better Reykjavik enable residents to propose, discuss, and vote on local initiatives, potentially broadening participation to those unable to attend in-person events \cite{barandiaran2024technopolitical,davies2020online,polis2015home}. At the same time, studies reveal that online platforms risk replicating offline inequalities. Groups with lower digital literacy or limited free time may remain underrepresented, and those who do participate might not have the means to navigate complex interfaces \cite{barendregt2024public,davies2020online,li2020analysis}.

Against this backdrop, generative AI, especially text-to-image models, has drawn attention as a way to lower the threshold for design conversations \cite{kucevic2024promptathon,vonbrackel2024equipping}. In Hamburg, for instance, the “Prompt-a-thon” format empowered residents to create futuristic visions of city districts using ChatGPT and DALL·E~3 \cite{vonbrackel2024equipping}. Researchers have documented potential advantages, including real-time iteration and the capacity to accommodate a wide spectrum of creative inputs \cite{guridi2024fake}. However, limitations persist, such as model biases in depicting marginalized communities and a tendency to produce polished imagery that may obscure feasibility \cite{kucevic2024promptathon}.

Recent scholarship has begun to examine the broader sociotechnical implications of generative AI in participatory urban design, emphasizing its entanglement with notions of citizenship, access, and visual culture. \citet{herrie2024democratization} introduce the concept of “visual citizenship” to describe how individuals engage with public issues through image-making practices enabled by generative AI. Drawing on four collaborative workshops in domains such as education and media, the authors distinguish between “formative” and “generative” modes of image-based participation, demonstrating that access to tools, skills, and interpretive authority significantly shapes the inclusivity of such engagements. This line of inquiry foregrounds the aesthetics of participation—not merely the output quality of AI images but the conditions under which image-making becomes a meaningful civic practice. Their findings point to the need for co-creative infrastructures that situate generative AI not only as a tool for representation but also as a processual space of negotiation and visibility.

In parallel, other studies have explored the operational potential and limitations of generative models in real-world co-design processes. \citet{guridi2024fake} conducted workshops and interviews in Los Angeles using image-generative AI tools to facilitate participatory park design, emphasizing how AI-mediated visuals can foster situated dialogue among diverse stakeholders. Their findings highlight both opportunities, such as improving conversational depth and spatial awareness, and concerns about bias, facilitation asymmetries, and misaligned expectations between designers and participants. \citet{valenca2025creating} similarly argue that emerging tools like UrbanistAI and Laneform address practical constraints of traditional design software by enabling real-time, user-modifiable visualizations of street layouts. However, they caution that the broader implications of such tools for creativity, contextual adaptability, and regulatory alignment remain underexplored. Together, these contributions call for sustained critical engagement with the ways generative AI tools reshape participatory design practices, both as instruments of expression and as boundary objects that mediate stakeholder relations and urban imaginaries.

\subsection{Inclusivity and barriers}
Inclusivity is a central theme in public engagement literature, emphasizing the importance of involving individuals with diverse abilities, backgrounds, and languages \cite{aitamurto2017crowdsourcing}. Achieving inclusivity demands careful planning around venue accessibility, scheduling flexibility, and mode of communication \cite{fors2021striving}. Digital solutions can expand reach, but they may also introduce new inequalities if the technology is not adapted to local contexts \cite{arana2021citizen,valenca2025creating}.

Bilingual or multilingual urban environments add layers of complexity. In Montreal, many residents naturally function in French, while advanced AI systems rely more heavily on English. This mismatch can result in misunderstandings or missed opportunities to incorporate nuanced local knowledge \cite{li2020analysis}. Observers warn that if AI-based tools do not handle multilingual prompts effectively, they risk creating an uneven playing field where only those comfortable with the dominant language can fully participate \cite{aitamurto2017crowdsourcing,davies2020online}.

The literature suggests that digital innovation holds the potential to reconfigure community consultations, making them more iterative, accessible, and visually oriented. Nonetheless, biases in AI systems, uneven digital literacy, and challenges of linguistic or cultural specificity are ongoing concerns. This study positions itself within this evolving domain by using Stable Diffusion XL to visualize citizen-generated prompts, potentially enabling a more inclusive process by lowering skill barriers. The methodology below describes how we operationalized this approach in a Montreal-based workshop and subsequent interviews, capturing both the strengths and limitations of generative AI for inclusive urban design.

\section{Methodology}
The study adopted a mixed qualitative approach, drawing on multiple data sources:
\begin{itemize}[leftmargin=*]
    \item Semi-structured interviews with six urban planning experts.
    \item An in-person workshop with 14 citizens, 5 computer scientists, 5 urban architects, and 5 note takers, who generated scenario-based prompts and visuals.
    \item Detailed analyses of workshop transcripts, participant-generated prompts and visualizations, and textual records of the expert interviews.
\end{itemize}

The objective was to understand how Stable Diffusion–generated visuals influence engagement, creativity, and representational accuracy in inclusive urban design consultations. Stable Diffusion was selected over other text-to-image models due to its open-source accessibility, customization potential, and ability to operate locally without proprietary constraints, making it better suited for participatory and community-driven contexts. After data collection, we employed various qualitative and computational techniques, including Latent Dirichlet Allocation (LDA) on interview transcripts \cite{blei2003lda}, LDA on workshop notes and transcripts, LDA on the textual prompts, and a word cloud analysis of the prompt corpus.

\subsection{Participant outreach}
\begin{figure}[t]
  \centering
  \includegraphics[width=\linewidth]{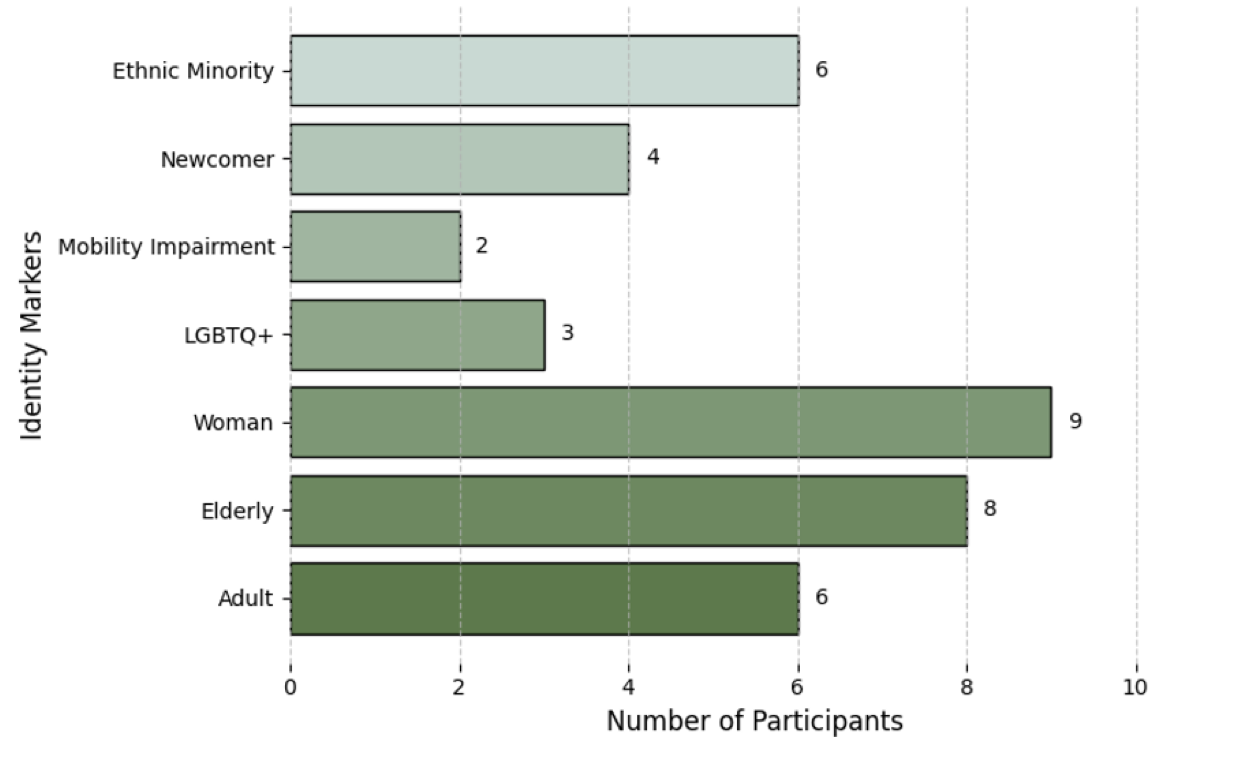}
  \caption{Participant Identities. This graphic depicts the self-reported identity markers of the 14 participants. Each cell corresponds to a demographic or identity attribute, visualizing the distribution of attributes among the workshop cohort.}
  \Description{Matrix-style infographic showing identity attributes of 14 workshop participants across age, mobility, cultural background, and language.}
  \label{fig:identities}
\end{figure}

We conducted an extensive outreach campaign to over 100 community and civic organizations in Montreal, combining email announcements and follow-up phone calls. We sought to ensure a broad mix of ages, abilities, and cultural affiliations. Four local architecture and planning firms also expressed interest, thereby providing professional insight. This process ultimately generated a workshop cohort of 14 participants who self-identified across various demographic markers. People from 2SLGBTQ+ communities, people using wheelchairs, elderly participants, and racialized and religious minorities. Linguistic diversity was notably present, with French-dominant, English-dominant, and bilingual individuals participating. Figure~\ref{fig:identities} displays a diagram of participant identity markers, highlighting the variety of demographic characteristics.

% Figure 1

\subsection{Expert interviews}
We conducted six semi-structured interviews with professionals in municipal facilitation, architectural and urban design, and parapublic planning. In addition to the core research questions, discussions explored:
\begin{itemize}[leftmargin=*]
    \item Current consultation and visualization practices.
    \item Potential benefits and risks of using generative AI in community engagements.
    \item Ethical concerns around biases, data usage, and the presentation of AI-generated images.
\end{itemize}
Each interview lasted approximately one hour. Transcripts were prepared for subsequent textual analyses, including LDA to identify latent themes related to AI uptake, inclusivity, or skepticism toward untested technologies. Saturation was assessed throughout the process; repetitive themes appeared after the fifth interview, and six interviews were deemed sufficient to capture the range of relevant perspectives.

\subsection{Workshop}
\begin{figure}[t]
  \centering
  \includegraphics[width=\linewidth]{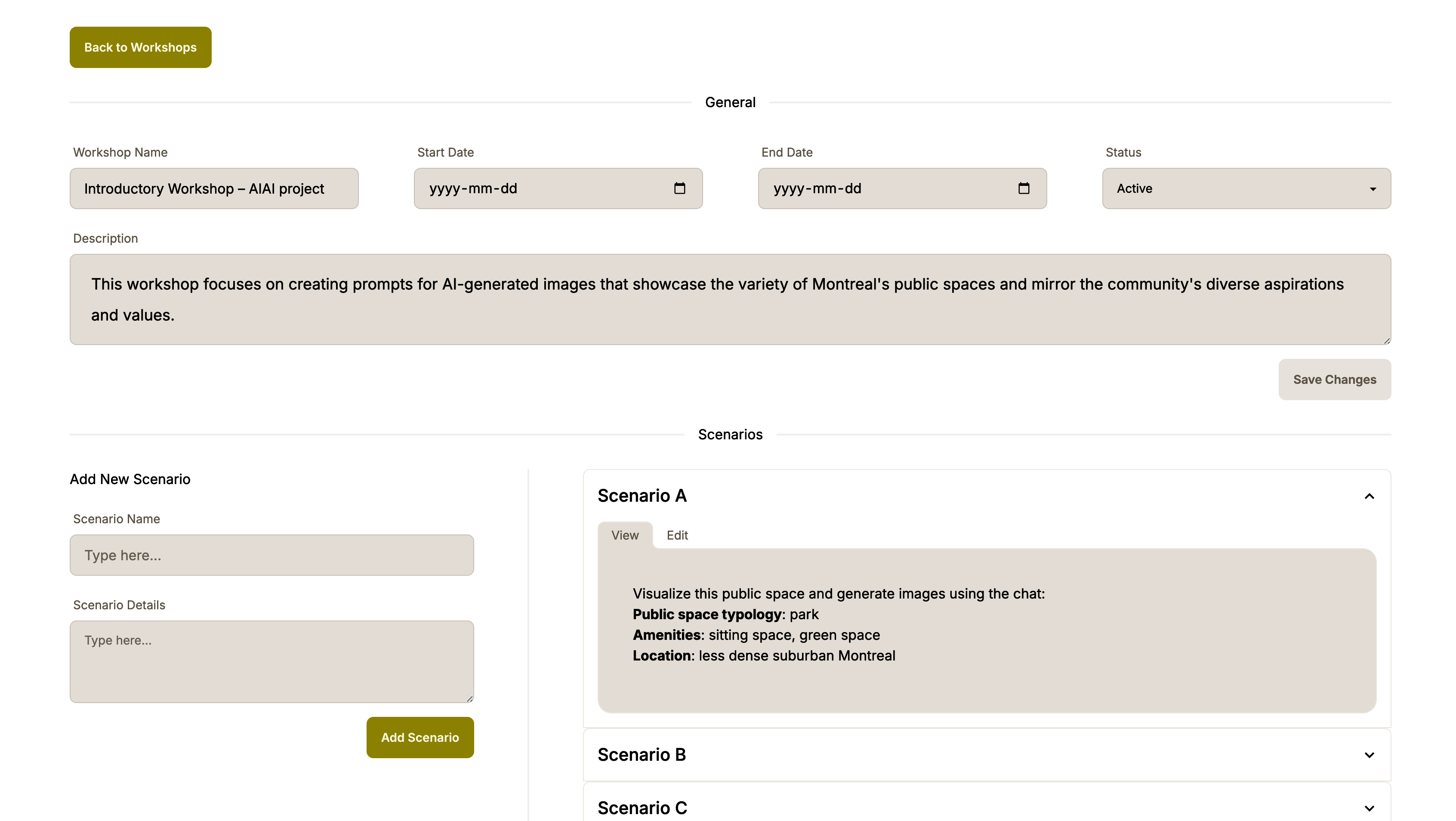}
  \caption{Platform Interface for WeDesign. This representation illustrates how participants create scenario descriptions and workshops. Each session stores multiple prompts and corresponding images, allowing iterative refinement.}
  \Description{Screenshot of WeDesign interface with fields for scenario description, prompt input, and a gallery of generated images for iterative refinement.}
  \label{fig:interface}
\end{figure}

The workshop took place in a focus group format. To facilitate better interaction with the AI model, we developed a platform called WeDesign. Participants were grouped into five teams, each including one AI specialist, one urban professional (architect or planner), and two or three citizens with varied backgrounds. WeDesign was introduced as a platform that uses Stable Diffusion XL to produce images based on textual prompts. Figures~\ref{fig:interface} and~\ref{fig:scenarioA} illustrate the interface, showing fields for scenario descriptions, textual prompts, and resulting image previews.

The groups engaged with five main scenarios: suburban parks (Scenario A), pedestrian promenades in a historical neighborhood (Scenario B), residential street spaces (Scenario C), dense downtown plazas (Scenario D), and a dense urban park near the waterfront (Scenario E). Optional scenarios, such as urban gardens or transit plazas, were available if time permitted. Each group spent approximately 30 minutes brainstorming, followed by a total of three hours (separated by a lunch break) typing prompts and iteratively refining them through discussions and real-time explorations. The objective was to capture community preferences related to elements such as ramps, diverse seating, lighting features, and local architectural characteristics.

Participants collectively generated about 440 textual prompts during the session. Each prompt was visualized through four images with varying seeds. They experimented with negative prompts to exclude overhead views or futuristic styles, discovering how to control the output’s level of detail and cultural cues. Citizens typically led the content decisions, while the AI specialist offered guidance on prompt structure. Urban professionals provided commentary on feasibility issues, though no final designs were expected. Participants expressed their preferences, and as the activity continued, the level of engagement grew.

% Figure 2 (interface)

\subsection{Data analysis}
All workshop discussions were audio-recorded, with field notes taken to capture participant dialogues as well as moments of confusion or enthusiasm. The WeDesign platform automatically archived each prompt generated during the sessions, along with metadata indicating which group submitted it and any user reactions, such as a ``heart'' symbol to signify preference. In addition, six interview sessions with urban experts were transcribed in full for subsequent analysis.

\paragraph{Qualitative coding and topic modeling.}
We conducted open coding on interview and workshop transcripts \cite{creswell2022research}, iteratively refining a codebook centered on inclusivity, language frictions, AI limitations, platform needs, and facilitation strategies. To triangulate, we applied LDA \cite{blei2003lda} to interviews and workshop notes after standard preprocessing (lowercasing, tokenization, stop-word removal in English and French, bigram detection). We explored $k \in \{5,\ldots,12\}$ and selected $k$ via topic coherence and interpretability checks by two authors. Given the small corpus, we use LDA strictly as a \emph{descriptive lens} to surface co-occurring terms rather than as a definitive model; all topic labels and interpretations were validated against coded excerpts. Two coders independently labeled 20\% of the data (stratified across groups); Cohen’s $\kappa$ was computed per top-level code and discussed until consensus; the remaining data were coded by one author with weekly peer debriefs.

\paragraph{Descriptive engagement metrics.}
To avoid purely impressionistic claims, we computed basic descriptors from platform logs: number of prompts submitted per group and per role (resident, AI specialist, urban professional), total images generated (4 per prompt), and the distribution of “heart” reactions per image. We also extracted speaking-turn counts from workshop transcripts.\footnote{We do not report inferential statistics, given the exploratory design; our intent is descriptive.} These descriptors contextualize qualitative themes on engagement and ownership (e.g., whether residents initiated most prompts, whether “hearts” clustered by group, and whether speaking turns balanced across roles).

\paragraph{Word cloud analysis of prompts.}
A separate LDA analysis was conducted on the 440 textual prompts generated by participants. We examined frequently used terms and their co-occurrence patterns, noting how certain topics (e.g., ``diverse user groups,'' ``cultural references,'' ``futuristic features'') recurred across scenarios. We also created a word cloud visualization to highlight the most commonly used words, which offered a quick overview of participants’ primary concerns—such as ``mobility,'' ``adaptive,'' ``greenery,'' ``accessibility,'' ``friendly,'' ``streets,'' and so on. This word cloud was cross-referenced with participant feedback, allowing us to see how rhetorical emphasis on inclusivity or accessibility emerged in textual form.

\paragraph{Synthesis and interpretation.}
We triangulated LDA topics with the open-coded categories, searching for convergences and divergences \cite{creswell2022research}. For example, if LDA extracted a topic about ``language friction'' from the workshop transcripts, we revisited those segments to analyze how participants described the bilingual environment. This approach helped corroborate or refine the themes that emerged from manual coding \cite{denzin2011handbook}. In certain cases, LDA-based topics confirmed frequent references to ``wheelchairs,'' ``ramps,'' and ``children’s play areas,'' which aligned with manual observations of participants’ inclusivity priorities.

\paragraph{Ethics and data governance.}
The study received Institutional Review Board approval. All participants provided informed consent for audio recording and the archival of prompts and images; resident participants received honoraria, and venues were wheelchair-accessible with bilingual facilitation. We communicated that AI outputs were conceptual and non-binding. We retain rights to prompts and generated images for research; upon acceptance, we will release a de-identified prompt log and model settings (without images depicting identifiable individuals), consistent with community norms and local policies \cite{mushkani2025urbanaigovernanceembed}.

\section{Results}

\subsection{Group dynamics}
Figure~\ref{fig:identities} (the identity diagram) shows that participants encompassed multiple age brackets, various cultural communities, and at least two individuals with a physical mobility impairment. Several participants had prior exposure to urban planning processes, but most were unfamiliar with generative AI. The bilingual mix (discussion in French-dominant, prompts in English-dominant) shaped how prompts were constructed. Each group contained an AI specialist to guide prompt syntax and an urban professional to provide domain context. Citizens contributed the majority of design content ideas.

During scenario discussions, participants alternated between their expressed keywords to be typed in the prompts. This cyclical arrangement gradually exposed all members to the platform’s interface and the nuances of Stable Diffusion–based text-to-image generation. Workshop notes indicate that group members often had to negotiate which features were most critical to include (e.g., accessibility vs.\ seating capacity vs.\ landscaping details), revealing the typical complexities of participatory design processes \cite{brabham2009crowdsourcing}.

\subsection{Prompt refinement}
The iterative nature of prompt refinement emerged as a key driver of engagement. Early prompts that attempted to bundle numerous requirements (e.g., accessible sidewalks, multi-lingual signage, diverse ethnicities, wheelchair users, pet areas) often generated cluttered or surreal images. Groups discovered that adding negative prompts or carefully ordering keywords produced more coherent outputs.

Participants also recognized that the Stable Diffusion model is not omniscient, frequently requiring repeated mentions of ``diverse people'' or ``universal design.'' One participant explained that ``we typed `wheelchair-friendly benches' but the image just created random wheelchairs everywhere.'' This underscores a mismatch between real accessibility needs and the model’s interpretation of prompts, corroborating critiques about generative AI’s difficulty in capturing nuanced features \cite{bendel2023image}.

\subsection{Realism versus surreal outputs}
Several participants wished to create realistic depictions of Montreal neighborhoods, specifying neighborhood names or famous buildings or suburban bungalows. However, the AI would sometimes produce stylized or vaguely Montreal or European cityscapes. This mismatch frustrated citizens hoping for direct references to local identity. The tension also prompted some participants to exploit the model’s imaginative capacity, requesting ``futuristic design,'' ``mushroom seating,'' or ``floating walkways,'' partially as a playful attempt to stretch the design conversation beyond conventional norms \cite{Mushkani2025JUM_MontrealStreets}.

Urban professionals occasionally cautioned against conflating these visually appealing scenes with feasibility. This phenomenon, where polished generative outputs overshadow practical constraints, aligns with findings that generative AI can create illusions of plausibility \cite{davies2020online,vonbrackel2024equipping}. The LDA analysis on workshop transcripts highlighted frequent co-occurrence of terms like ``Montreal,'' ``futuristic,'' and ``imaginative'' alongside references to ``limitations'' or ``reality,'' reflecting how participants themselves recognized this duality.

\subsection{Inclusivity and local culture}
\begin{figure}[t]
  \centering
  \includegraphics[width=\linewidth]{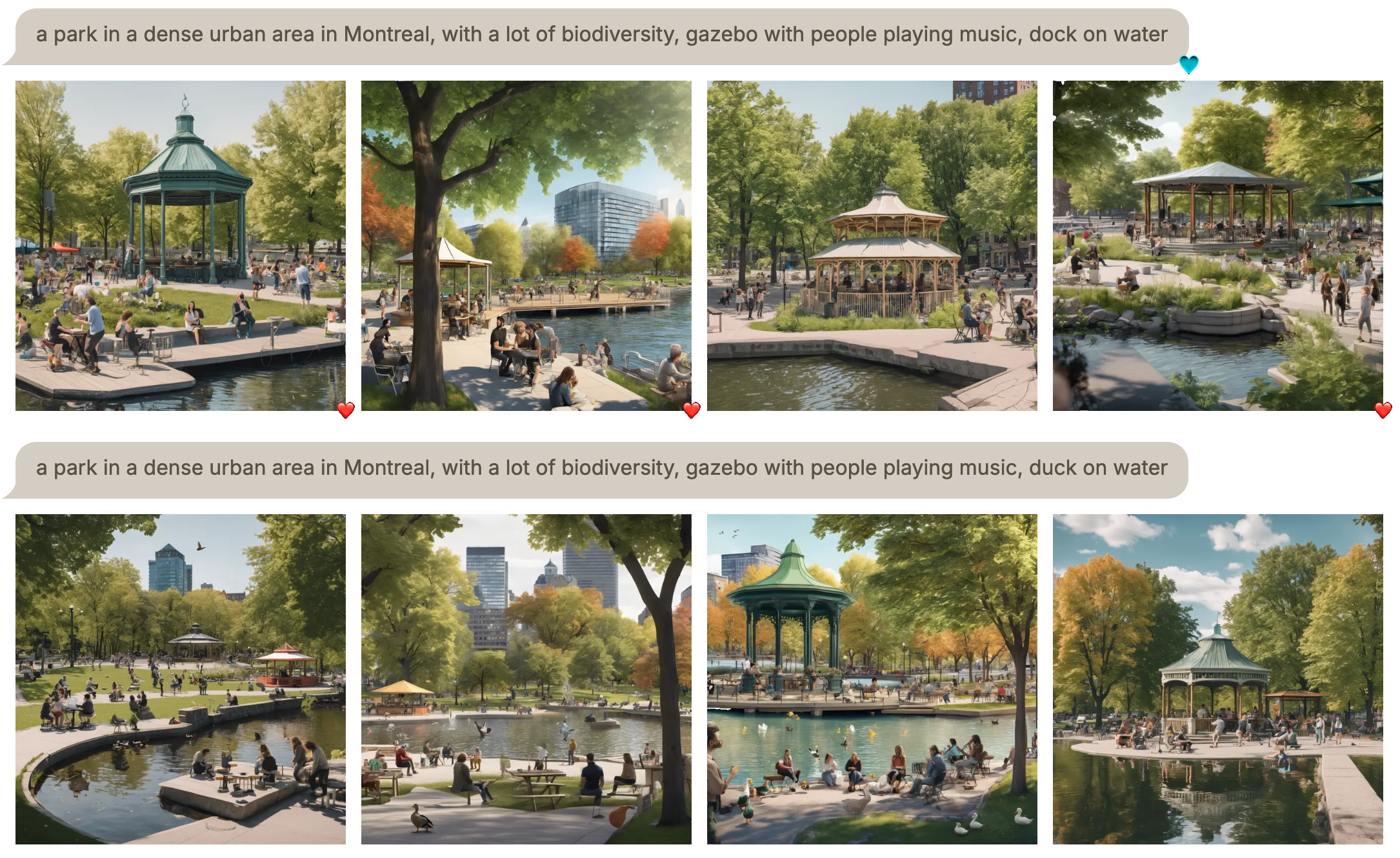}
  \caption{Park with waterfront, biodiversity, and community engagement. The prompt asked for a park in a dense urban context, highlighting rest areas, water edges, interactive zones, and accessible paths.}
  \Description{AI-generated urban waterfront park with trees, seating, open spaces, and water edge; some inclusive elements missing.}
  \label{fig:scenarioE}
\end{figure}

\begin{figure}[t]
  \centering
  \includegraphics[width=\linewidth]{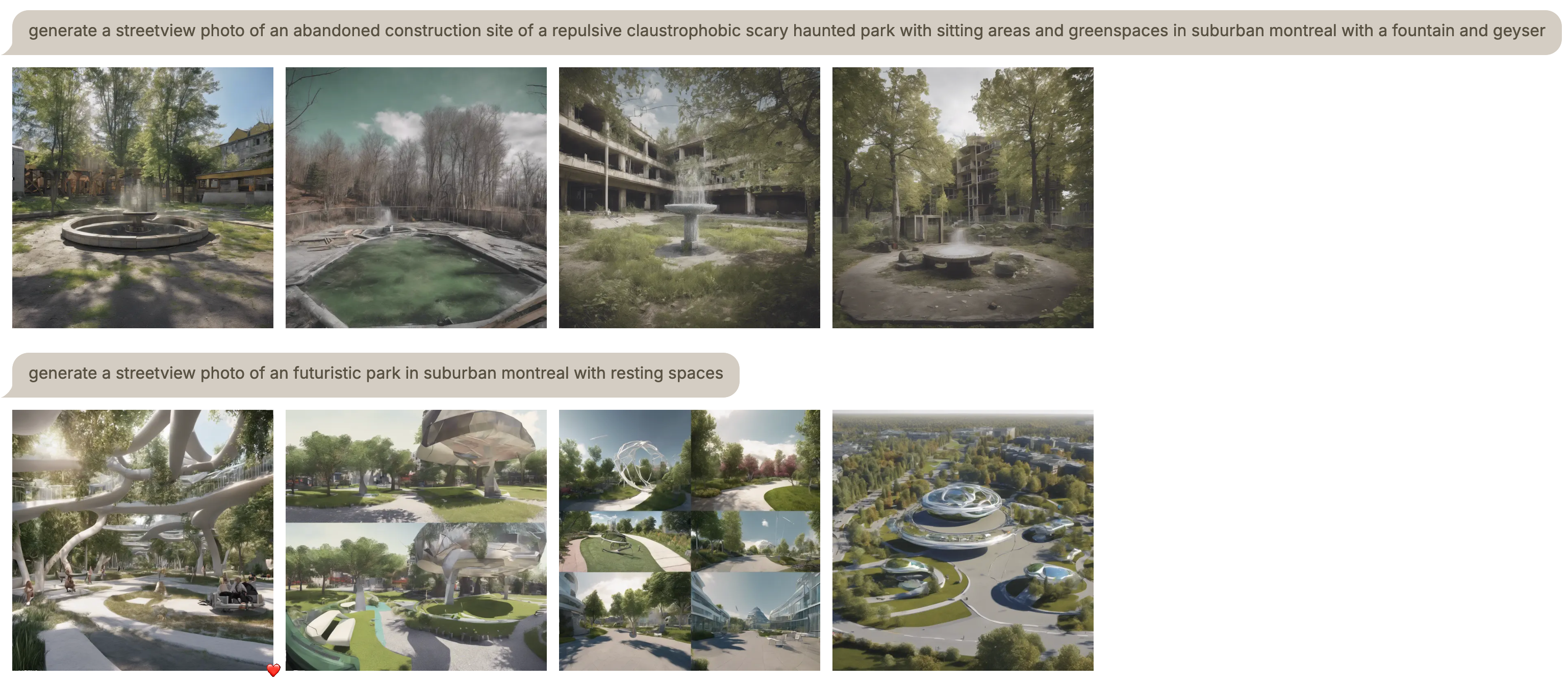}
  \caption{Suburban Montreal public spaces. Participants explored negative prompting and futuristic ideas, with preferences indicated using a heart symbol.}
  \Description{Grid of AI-generated suburban public space variants; one image marked with a heart as preferred.}
  \label{fig:scenarioA2}
\end{figure}

\begin{figure}[t]
  \centering
  \includegraphics[width=\linewidth]{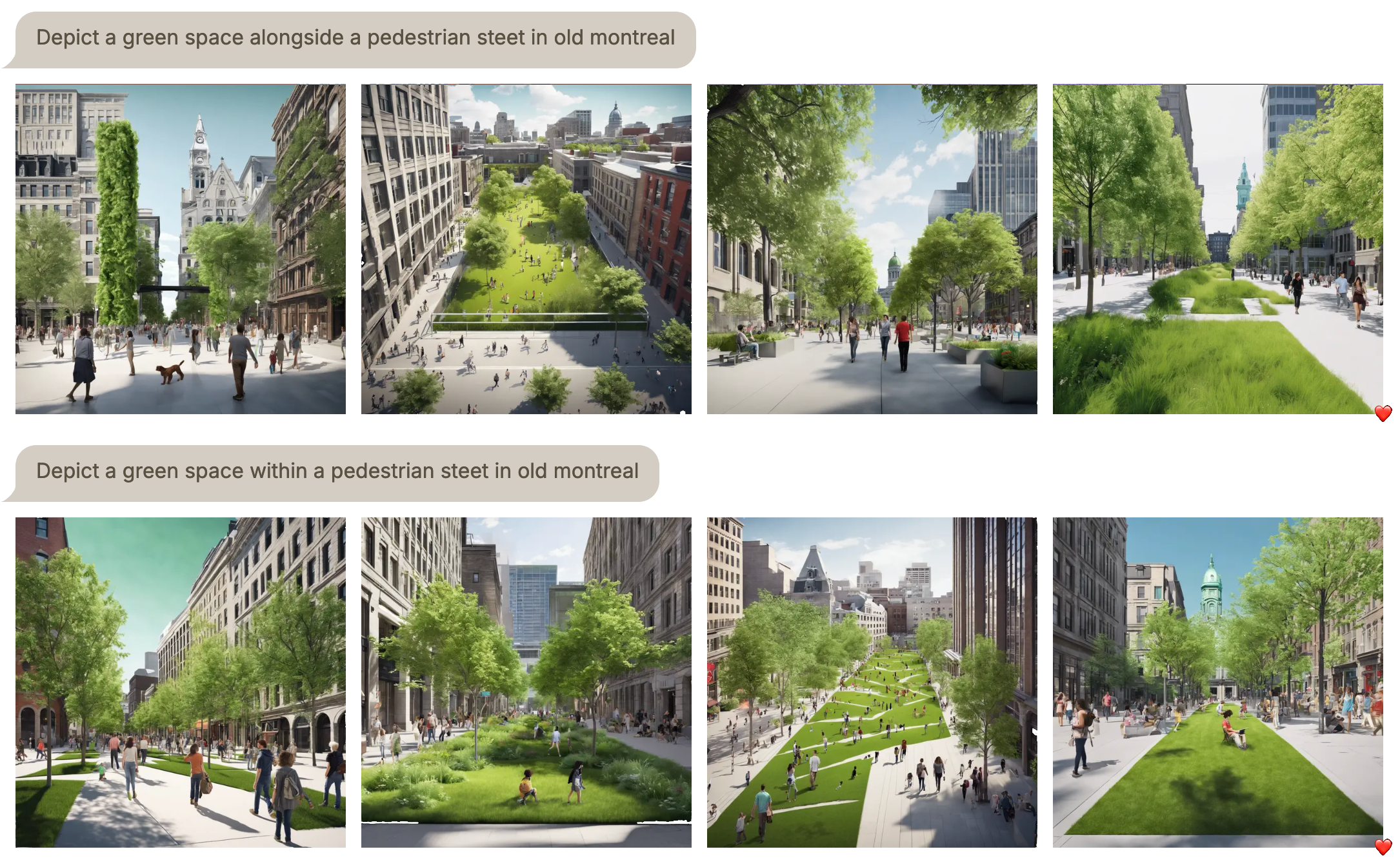}
  \caption{Pedestrian promenade in a historical Montreal neighborhood. Participants noted that it balanced heritage and modern amenities.}
  \Description{AI-generated lively pedestrian street with historic facades and modern street furniture.}
  \label{fig:scenarioB}
\end{figure}

Groups consistently sought images that depicted spaces accommodating diverse ethnic backgrounds, varied age groups, and individuals with mobility constraints \cite{MushkaniKoseki2025Habitat_StreetReview}. Yet repeated references to these themes did not reliably yield authentically diverse visuals. Rather, the AI occasionally inserted tokenistic elements (e.g., symbolic flags, depictions of mobility devices) without integrating them organically into the scene. Figures~\ref{fig:scenarioA}--\ref{fig:scenarioB} provide snapshots of typical results:

% Figure 3 (Scenario A)
\begin{figure}[t]
  \centering
  \includegraphics[width=\linewidth]{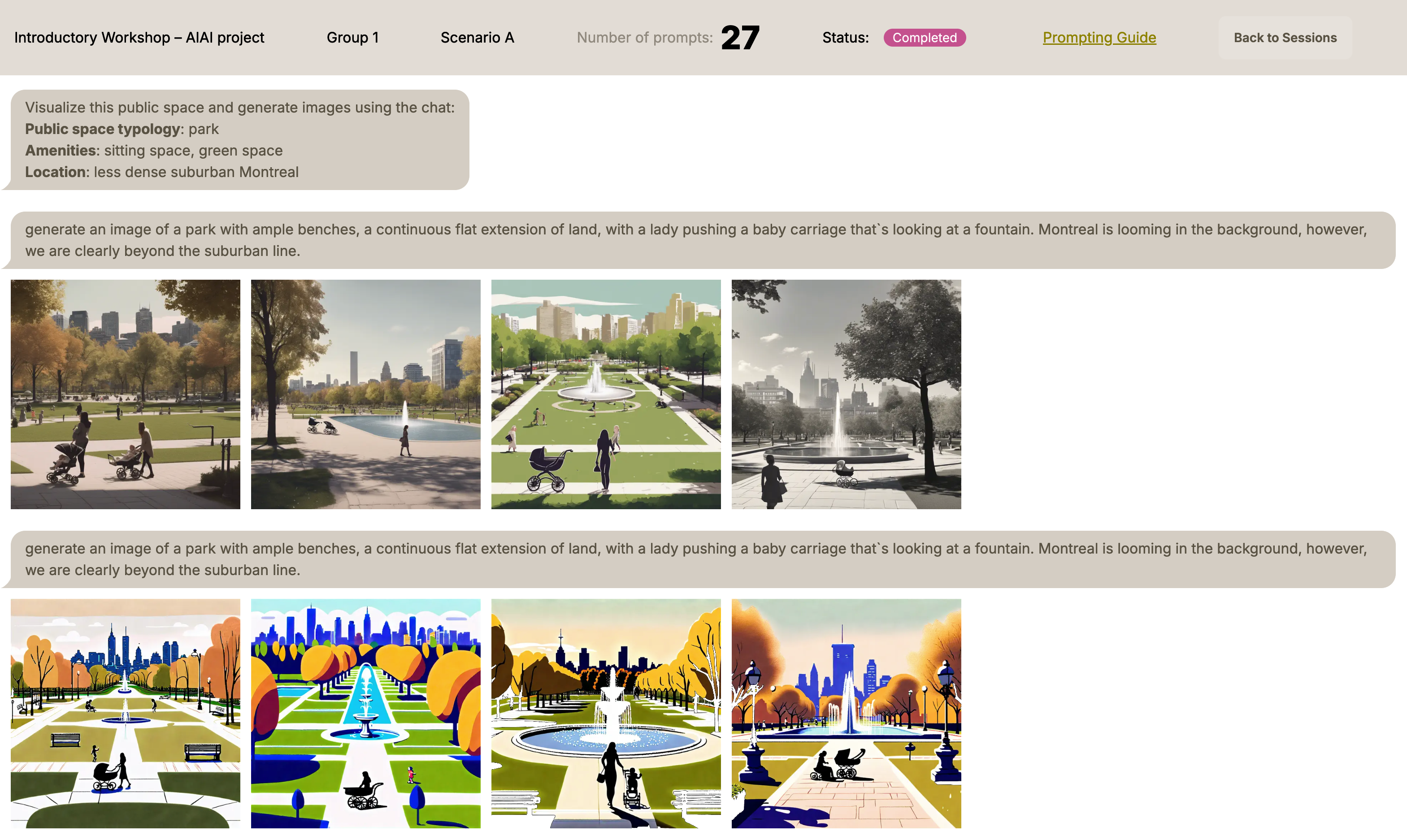}
  \caption{Suburban park in a less dense Montreal setting. A description of Scenario A, followed by an initial prompt, and then the same initial prompt with ``realistic'' added as a negative keyword. The prompt focused on sitting areas, family-friendly features, and a sense of calm.}
  \Description{AI-generated images of a suburban park emphasizing greenery and benches; panels compare prompt variants for Scenario A.}
  \label{fig:scenarioA}
\end{figure}

% Figure 4 (Scenario E)

% Figure 5 (Scenario A2)

% Figure 6 (Scenario B)

LDA on workshop transcripts revealed a topic dominated by terms such as ``wheelchair,'' ``ethnic,'' ``diversity,'' ``accessibility,'' ``user,'' ``spaces,'' and ``ignored.'' This suggests that participants often discussed the mismatch between their inclusive ambitions and the AI’s outputs, acknowledging the structural constraints inherent in training data \cite{bendel2023image}. Workshop notes further show that attempts to depict users with reduced‑mobility needs rarely yielded functionally accurate arrangements. While the model inserted wheelchairs, it seldom adjusted pavement gradients, ramp slopes, or bench clearances \cite{mushkani2025streetreviewparticipatoryaibased}. Participants traced this to training data sparsity and to the prompt interface, which did not allow region‑specific in‑painting to correct single elements without regenerating the full scene.

\subsection{Language barriers}
\citet{davies2020online} underscore the importance of language adaptation in digital platforms. In this workshop, French-speaking participants tried prompts in English, because Stable Diffusion XL tended to perform better with English prompts, producing more accurate or contextually relevant images. The LDA topic modeling of workshop notes found a cluster referencing ``language,'' ``French,'' ``English,'' ``translation,'' ``confusion,'' and ``culture.'' This cluster aligned with discussion transcripts in which participants debated the practicality of simultaneously conversing in French but prompting in English. The group-level data show repeated attempts to approximate local references, sometimes settling for simpler English words to achieve comprehensible results. Participants therefore requested automated prompt translation or the embedding of large language models into the interface to reduce repeated manual switching \cite{Mushkani2025ICML_RightToAI}.

\subsection{Creativity and engagement}
While participants encountered frustrations with representational gaps, they also described the workshop as more engaging and less ``dry'' than conventional presentations. Immediate visual outputs and the iterative prompt approach encouraged deeper discussion, with one participant stating that ``it feels like we’re all brainstorming together, seeing what works or doesn’t, instead of just listening to a lecture.'' The LDA topics on workshop transcripts frequently included co-occurrences such as ``fun,'' ``try again,'' ``prompt tweak,'' and ``discuss,'' underscoring the interactive energy.

% Figure 7 (Word cloud)
\begin{figure}[t]
  \centering
  \includegraphics[width=\linewidth]{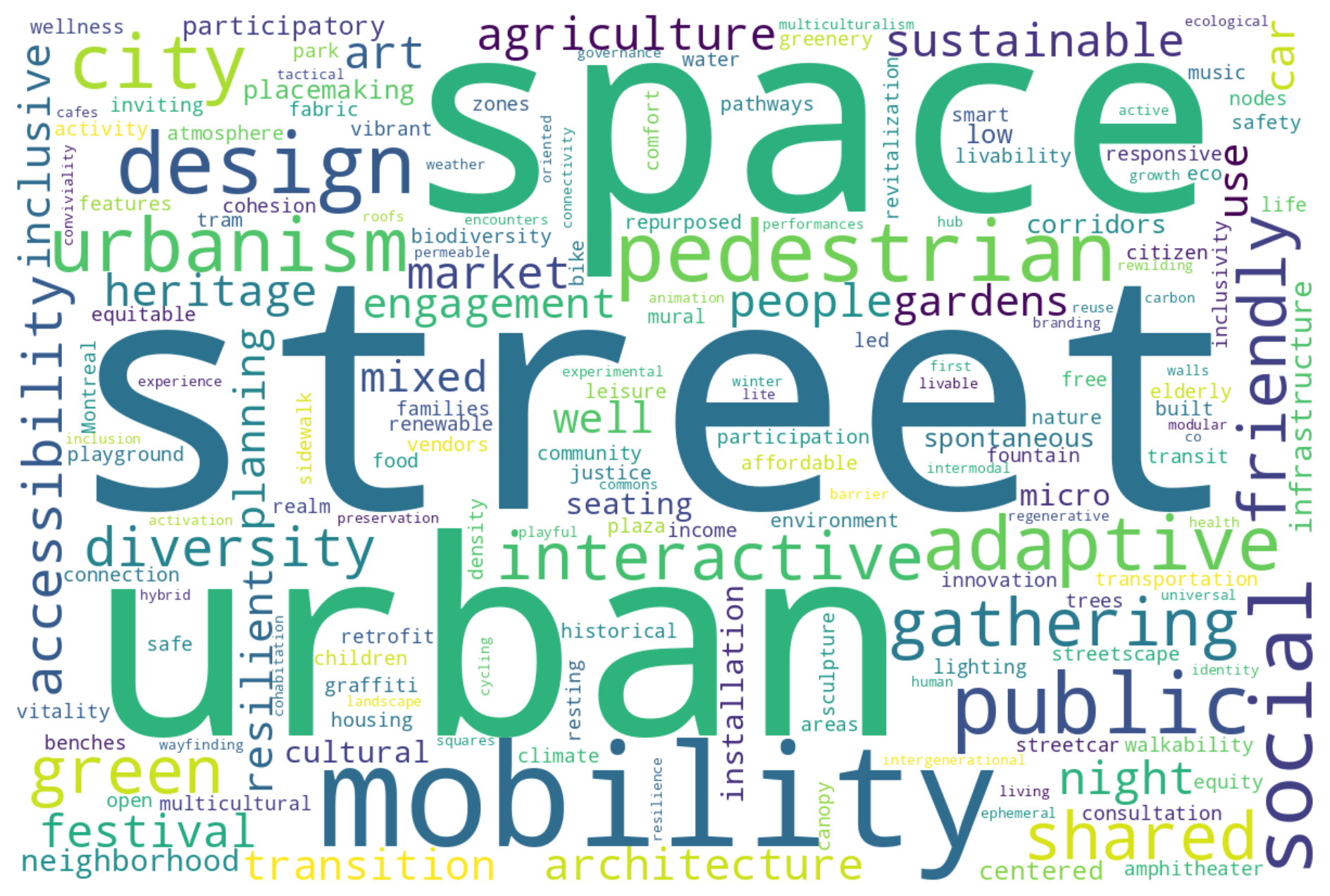}
  \caption{Word cloud of frequently mentioned themes in participant prompts (e.g., accessibility, mobility, greenery, diversity).}
  \Description{Word cloud highlighting terms such as accessibility, mobility, greenery, diversity, pedestrian, and community.}
  \label{fig:wordcloud}
\end{figure}

An additional dimension of creativity emerged when participants encountered unexpected elements in the generated images. For instance, even if a prompt did not specify the use of wood, the sudden appearance of a wooden bench in the visualization often inspired new directions. Participants would then explicitly incorporate ``wooden bench'' into subsequent prompts, demonstrating how serendipitous outputs could broaden their imaginative scope and iterative design thinking.

Simultaneously, the word cloud created from the 440 prompts (Figure~\ref{fig:wordcloud}) highlighted key terms like ``street,'' ``space,'' ``diversity,'' ``mobility,'' ``adaptive,'' ``green,'' ``pedestrian,'' ``accessibility,'' and ``friendly.'' This visual showed that participants consistently emphasized community-oriented and inclusive features, a pattern corroborated by the manual coding of transcripts. The synergy of LDA findings, word cloud highlights, and direct quotes indicates that the generative process amplified participants’ focus on user diversity, accessible amenities, and local context, even if some of the results were imperfectly realized in the Stable Diffusion images.

\subsection{Expert interviews}
Semi‑structured interviews with six practitioners generated three cross‑cutting themes that correspond to the study’s research questions.

\paragraph{Role of visualization.}
All experts emphasized that citizen engagement is embedded contractually in large-scale developments in Montreal. As two architects noted, ``Consultation is mandatory on most large-scale projects, part of the contract since they’re developed at such big scales'' (Experts C and F). Visualization tools were described as central both to divergence, by exploring a range of options before converging on common ground, and to convergence, by testing which options resonate with different groups (Experts A and B). Experts reported that the early introduction of visuals helps non-specialist stakeholders articulate preferences, thereby enhancing creative exchanges and a sense of ownership over emerging concepts (Experts A to F).

\paragraph{Challenges to inclusive representation.}
Practitioners identified logistical, structural, and methodological barriers to reaching representative samples. Scheduling sessions at times accessible to caregivers, providing honoraria, and selecting appropriately sized focus groups were cited as necessary but resource‑intensive measures: ``It’s a challenge to draw what you think – instead if you can give input on the image, people feel they’re adequately representing what they see in their mind'' (Expert C). Recruitment often relies on third‑party firms to avoid bias, yet funding limits may curtail the number or duration of workshops, potentially skewing participation toward those with greater availability or digital literacy (Experts A to C, and F).

\paragraph{Constraints of generative AI.}
Experts expressed cautious interest in real-time image generation. They envisioned AI tools that could ``represent ideas we’re talking about in the moment,'' while flagging the risk that ``an infinity of images can create expectations too high'' (Expert E). Key requirements included context-specific cues, open-source transparency to allow community review of training data, and mechanisms for iterative refinement with participant feedback (Experts A and D). A professional remarked that they were ``skeptical about the universality of inclusiveness,'' underlining concerns about using generative models trained on non-local datasets (Expert C). Ethical considerations centered on data provenance, bias mitigation, and clear communication that AI outputs are conceptual rather than contractual.

Collectively, these insights align with the workshop findings: visuals drive engagement and iterative creativity, yet current AI outputs struggle to convey nuanced local and inclusive features. Overall, the expert interviews elucidated the strategic potential of generative AI in participatory design, provided it is used judiciously and transparently. Professionals suggested incorporating generative AI as an exploratory tool early in the consultation process, with subsequent iterations refined through direct community feedback. They also recommended further pilot studies and policy-level changes to better accommodate and enhance participation through the use of text-to-image models. This approach aligns with their broader recommendation for iterative and adaptive engagement strategies that reflect authentic community insights rather than fixed design solutions.

\subsection{Platform features requested}
Across focus groups and interviews, participants converged on four functional requirements for future iterations of WeDesign. First, they requested region-specific in-painting, such as a brush-based tool to modify discrete image areas without discarding satisfactory background elements. Second, they emphasized the need for multilingual support, including seamless prompt translation and synonym suggestions across French and English to preserve local nomenclature while leveraging the model’s stronger English parsing. Third, they proposed group-based voting, suggesting a built-in mechanism to rank or discard variants collectively rather than relying on ad hoc verbal consensus \cite{mushkani2025negotiativealignmentembracingdisagreement}. Fourth, they advocated for preference sliders or toggle buttons that would allow users to weight criteria such as accessibility, biodiversity, or heritage, providing non-technical users with an intuitive way to steer generation parameters. Participants stressed that these features should be delivered through an open-source codebase to enable public inspection and iterative community contributions.

\section{Discussion}
The results indicate that using text-to-image tools during consultations can amplify citizen voices and stimulate creativity in ways that differ from typical top-down consultations \cite{Mushkani2025AIES_CoProducingAI}. The study aligns with prior research suggesting that text-to-image tools help reduce the skill gap in illustrating design ideas \cite{vonbrackel2024equipping,kucevic2024promptathon}. Workshop participants found the process more collaborative, with iterative prompt generation serving as a catalyst for group reflection on specific design features \cite{li2020analysis}.

However, representational accuracy remains a significant obstacle \cite{Mushkani2025ICML_LIVS}. The system’s partial depiction of inclusivity, overshadowed by tokenistic or missing features, reflects broader concerns that generative AI replicates biased training data \cite{bendel2023image}. Additionally, bilingual friction limited some participants’ ability to express their ideas fully, reinforcing the literature’s emphasis on digital inclusivity in multilingual regions \cite{davies2020online}. The LDA topics extracted from transcripts and workshop notes confirmed that such tensions formed a major thread of conversation, emphasizing the need for local adaptation in AI models.

This interplay between imaginative freedom and practical constraints also mirrored typical consultative tensions \cite{guridi2024fake}. On one hand, the capacity to produce realistic, whimsical or idealistic images encouraged out-of-the-box thinking. On the other, repeated references to ``reality checks'' indicated concern that communities might develop unrealistic expectations. Expert interviews confirmed that final decisions depend on budget, zoning, and political will—factors rarely visible in generative images \cite{arana2021citizen}.

The LDA analysis of prompts likewise highlighted the recurring desire for benches, waterfront access, safe pathways, and user-friendly architecture, suggesting a consistent interest in communal well-being. Word cloud representations underscored themes such as greenery, diversity, and access, reinforcing that participants wanted images reflecting active, inclusive, socially cohesive spaces. Yet while Stable Diffusion–based AI can prototype those images quickly, bridging the gap from generic conceptual visuals to contextual visuals still remains elusive. Nevertheless, using text-to-image models during community consultations requires robust facilitation and transparent discussions about constraints \cite{davies2020online}.

\paragraph{Design Implications for Equitable AI-Mediated Consultations}
\begin{enumerate}
    \item \textbf{Region-specific in-painting as a core workflow.} Brush-based edits with constraints (e.g., sidewalk slope, bench clearance) let facilitators fix one accessibility flaw without regenerating a scene.
    \item \textbf{Multilingual prompt mediation.} Embed bilingual prompt translation, synonym expansion for local toponyms, and glossaries, with inline warnings for ambiguous terms.
    \item \textbf{Participatory preference steering.} Expose non-technical sliders (accessibility, biodiversity, heritage) that map to prompt templates and negative-prompts, logged for auditability.
    \item \textbf{Expectation management cues.} Watermark images (“conceptual, not final”), overlay feasibility hints, and provide side-by-side “as-is vs. concept” views.
    \item \textbf{Role-aware facilitation.} Make the \emph{resident} the prompt author-of-record; require experts to frame feasibility as constraints to be explored, not barriers to ideation.
    \item \textbf{Provenance and audit.} Log prompt, seed, sampler, steps, and guidance for each image; enable export as part of the public record.
\end{enumerate}

\section{Limitations and Reflexivity}
Our mixed-methods study is exploratory with a modest sample size and a single-city context. Resident participants were self-selected through outreach and may overrepresent individuals comfortable with workshops; we mitigated this through diverse scheduling and honoraria but cannot claim representativeness. Facilitation choices (e.g., “converse in preferred language, prompt in English”) likely shaped outcomes. To counter this, we documented facilitation moves and emphasized resident-authored prompts. Finally, generative images can inflate expectations; we watermarked outputs and framed them as conversation starters, not design approvals.

\section{Conclusion}
This study examined how WeDesign, anchored in Stable Diffusion XL, can facilitate a novel form of community consultation that aims to give citizens a more direct voice in shaping urban design ideas. By analyzing workshop transcripts, participant prompts, interviews, and applying LDA topic modeling and word cloud analyses, we identified several key findings:

\begin{itemize}[leftmargin=*]
    \item \textbf{Increased engagement:} Generative AI–driven visualizations can create a more collaborative and flexible environment, lowering the traditional skill barriers to design illustration. Participants appreciated the dynamic, iterative nature of the platform, describing it as more enjoyable than conventional consultations.
    \item \textbf{Challenges in depicting inclusivity and local detail:} Despite frequent references to local locations, diverse populations, and accessibility, the outputs often overlooked or tokenized these features. The system struggled to embed reduced-mobility design standards, and language biases hindered French-speaking participants, who obtained more generic results when prompting in French.
    \item \textbf{Balancing aspirational and realistic visions:} Many users experimented with futuristic or whimsical prompts, reflecting the creativity spurred by quick AI-based rendering. Yet, concerns arose around feasible implementation, echoing critiques that generative outputs can appear deceptively polished \cite{kucevic2024promptathon}.
    \item \textbf{Analytical insights from LDA and word clouds:} Topic modeling and word cloud analysis confirmed that participants repeatedly invoked themes of inclusivity, safety, and community identity. This pattern aligned with manual observations of workshop dialogues, illustrating that the textual corpus carried consistent priorities around user-centric and culturally sensitive design.
    \item \textbf{Platform design priorities:} Participants and experts independently called for open‑source development, region‑specific in‑painting, integrated bilingual support, collective image‑ranking tools, and slider‑based preference inputs to improve accuracy and transparency.
\end{itemize}

Looking ahead, participants saw potential for using text-to-image models in asynchronous or online consultations, expanding reach to individuals who cannot attend in-person events. Large language models could further support prompt translation, help refine prompts for local cultural references, and guide textual brainstorming. Such extensions may mitigate issues related to bilingual friction or limited AI understanding of local architecture.

Several concluding points emerged during the process. Citizens realized that Stable Diffusion XL does not automatically embed local knowledge, prompting frustration when visual outputs ignored certain details. Yet many participants valued the creativity and fun factor, recognizing it as a departure from more tedious consultation formats. The possibility of asynchronous sessions, where participants upload or refine prompts remotely and engage in voting or brainstorming, could improve accessibility, addressing constraints such as venue inaccessibility and scheduling conflicts. Indeed, this approach might reduce some logistical hurdles, consistent with calls to broaden digital civic engagement.

Future research might extend these findings by piloting WeDesign at a larger scale, incorporating more advanced features, or systematically evaluating whether repeated prompt iterations can yield more accurate diversity representation. Collaborations with municipal agencies could also integrate text-to-image–based outputs into official planning documents, potentially deepening iterative feedback loops between the public, experts, and decision-makers. Finally, addressing the ethical implications of data governance, including who retains rights to the generated images and how these might shape public perception, remains a priority.

\section{Platform availability}
An updated version of WeDesign+, integrating the requested features, will be released as open-source software with public access to the code and documentation.

% ----------------- References -----------------
\bibliography{references}

\end{document}